\documentclass[journal ,12]{IEEEtran}
\usepackage{colortbl}
\usepackage[table]{xcolor}

\usepackage{tikz}
\usepackage{url}
\usepackage{float} 
\usepackage{amsmath,amsfonts,amssymb,mathrsfs,amsthm,amsbsy,graphicx,color,bm,balance,mathtools,xfrac,nccmath,siunitx,cite,soul}
\usepackage{multirow}

\usepackage[font=scriptsize,labelfont=bf]{caption}

\usepackage{pifont}

\usepackage{pgfplots}
\pgfplotsset{compat=newest}
\usetikzlibrary{plotmarks}
\usetikzlibrary{arrows.meta}
\usepgfplotslibrary{patchplots}
\usepackage{grffile}
\usetikzlibrary{plotmarks}

\usepackage{algpseudocode}
\usepackage{algorithm}
\algnewcommand\algorithmicinitialize{\textbf{Initialize:}}
\algnewcommand\Initialize{\item[\algorithmicinitialize]}

\newcommand{\abc}{\text{AmBC}\xspace} 
 \usepackage[utf8]{inputenc}
\usepackage{booktabs, caption, makecell,xspace}

\usepackage{threeparttable}

 \theoremstyle{definition}
 \makeatletter
\newcommand{\printfnsymbol}[1]{%
  \textsuperscript{\@fnsymbol{#1}}%
}
\makeatother
\begin{document}
\title{Adversarial  Score-Based Generative Models for MMSE-achieving  \abc Channel Estimation }
\author{Fatemeh Rezaei, \IEEEmembership{Member, IEEE}, S. Mojtaba Marvasti-Zadeh, \IEEEmembership{Member, IEEE},   Chintha Tellambura, \IEEEmembership{Fellow, IEEE,}   Amine Maaref, \IEEEmembership{Senior Member, IEEE}
\thanks{F. Rezaei and C. Tellambura are with the Department of Electrical and Computer Engineering, University of Alberta, Edmonton, AB, T6G 1H9, Canada (e-mail: \{rezaeidi, ct4\}@ualberta.ca).  \\
\indent S. M. Marvasti-Zadeh is with the Departments of Computing Science and Renewable Resources, University of Alberta, Edmonton, AB, T6G 1H9, Canada (e-mail: seyedmoj@ualberta.ca).\\
 \indent A. Maaref is with Huawei Canada, 303 Terry Fox Drive, Suite 400, Ottawa, Ontario K2K 3J1 (e-mail: amine.maaref@huawei.com).}
\vspace{-10mm}}


\maketitle

\begin{abstract}
This letter presents a pioneering method that employs deep learning within a probabilistic framework for the joint estimation of both direct and cascaded channels in an ambient backscatter (\abc) network comprising multiple tags. In essence, we leverage an adversarial score-based generative model for training, enabling the acquisition of channel distributions. Subsequently, our channel estimation process involves sampling from the posterior distribution, facilitated by the annealed Langevin sampling technique. Notably, our method demonstrates substantial advancements over standard least square (LS) estimation techniques, achieving performance akin to that of the minimum mean square error (MMSE) estimator for the direct channel, and outperforming it for the cascaded channels. 
\end{abstract}

\vspace{-2mm}
\section{Introduction}
\IEEEPARstart{A}{mbient} backscatter communication (\abc) is an emerging enabler of passive Internet-of-Things (IoT) networks, where ultra-low-power backscatter tags, rely solely on modulating incident radio frequency (RF) signals for data communication \cite{Rezaei2023}. Tags have compact storage and limited capacity/power, necessitating continuous recharging of the batteries via energy harvesting (EH). 

Accurate channel estimation is essential for the reader to detect tag signals in \abc networks. However, these networks present unique challenges, including the limited processing capabilities of tags, the presence of weak tag signals, and mutual interference among multiple tags. Additionally, tags cannot inherently generate pilots; instead, they must reflect pilots from an external source. Estimating cascaded (dyadic) channels in such networks is particularly challenging.

\begin{figure}[!t]\centering \vspace{-0mm}
 \includegraphics[width=0.8\linewidth]{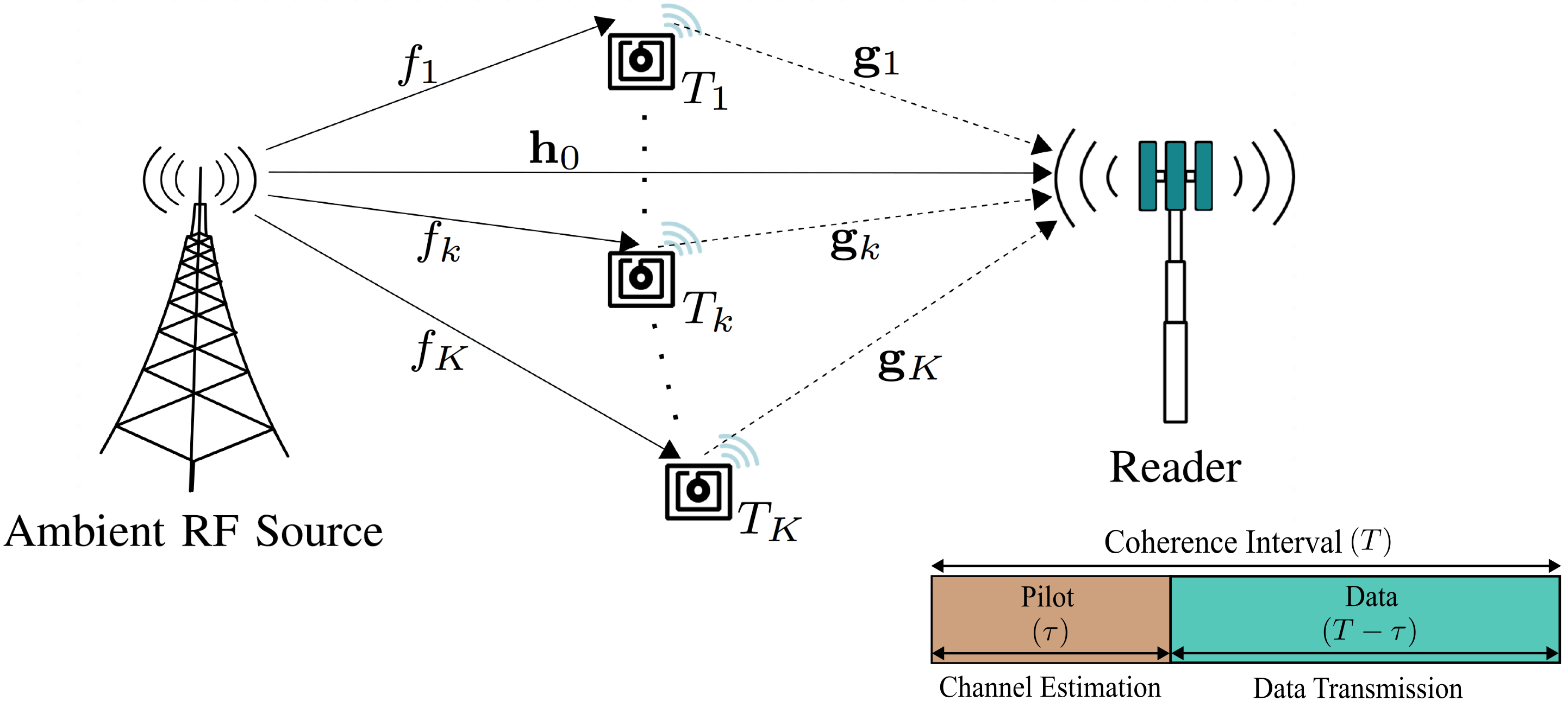}
\vspace{-1mm}
    \caption{\abc system and channel coherence interval.}
    \vspace{-6.5mm} \label{fig_SystemModel}
 \end{figure}

In a typical \abc system (Fig. \ref{fig_SystemModel}), the reader estimates channel state information (CSI) for two distinct types  of channels, namely: i) the direct channel stretching from the RF source to the reader ($\mathbf{h}_{0}$), and ii) the cascaded (or dyadic) channel $f_{k} \mathbf{g}_{k}$, which is from the RF source to the $k$-th tag and back to the reader. This dyadic channel exhibits distinct fading behaviors compared to conventional one-way wireless links, resulting in more pronounced fades \cite{griffin2010fading, Rezaei2023}.

The amalgamation of these two  channels at the reader compounds the complexity of obtaining separate estimates for each. Consequently, while classical methods and machine learning (ML) techniques have proven effective in conventional channel estimation scenarios \cite{biguesh2006training, Hu2021, Arvinte2023}, the aforementioned distinctive challenges of \abc hinder a straightforward application of these established approaches.
Nevertheless, the studies  \cite{Shuo2018, Zhao2019, Liu2021, Wang2022, rezaei2023timespread} investigate  \abc channel estimation via both classical and deep learning (DL) techniques. These works utilize pilot sequences sent by the RF source. Notably,  \cite{Wang2022} harnesses denoising blocks and exploits successive interference cancellations to derive estimates for the direct and cascaded channels. Similarly, \cite{Liu2021} employs convolutional neural networks to sequentially estimate the direct channel and cascaded channels. This work trains a deep residual network tailored to each tag's unique characteristics. However, works   \mbox{\cite{Shuo2018, Zhao2019, Liu2021, Wang2022}} estimate the cascaded channels indirectly by subtracting the direct channel estimate. This leads to error propagation and an increase in the mean square error (MSE). Moreover, each channel is estimated using a fraction of the pilot sequence. Hence, their efficiency diminishes as the number of tags increases.

In contrast, \mbox{\cite{rezaei2023timespread}} estimates  the cascaded channels  directly, avoiding error propagation. Besides, each channel estimate is computed over the entire pilot sequence, reducing the MSE even for shorter pilot sequences. As this will increase data transmission time, spectral efficiency will improve, and the total power consumption will reduce. Study \mbox{\cite{rezaei2023timespread}} develops  orthogonal pilot sequences, optimizing {\abc} channel estimation, surpassing prior arts  \mbox{\cite{Shuo2018, Zhao2019, Liu2021, Wang2022}}. Nevertheless, the classical minimum mean square error (MMSE) estimator \mbox{\cite[Section 11.4]{kay1993fundamentals}}, although superior to the least square (LS) estimator, requires channel correlation statistics and precise channel distribution \mbox{\cite{kay1993fundamentals}}, which is not available in general.

To tackle this, we use adversarial score-based generative models, which learn and approximate a dataset's probability distribution. They train a neural network to estimate the score function, learning the gradient of the log-density of the data distribution.  These models excel in implicit data density estimation, handling multi-modal distributions, sampling complex distributions, preventing mode collapse, aiding evaluation, and providing interpretable gradients \mbox{\cite{song2019generative, song2020improved, jolicoeur2020adversarial}}.

 However, prior to our study, they had never been explored for  \abc channel estimation.  To surmount the challenges of \abc channel estimation, and to achieve the optimal  MMSE estimator performance, we thus introduce an innovative adversarial score-based generative model.   It uniquely addresses the joint estimation of both direct and cascaded channels ($K>1$) -- Fig.~\ref{fig_SystemModel}. Yet, these two sets of channels display distinct fading behaviors, rendering precise data distribution modeling highly intricate. Our approach achieves accurate estimation of the channel probability distribution using the score function (defined as the gradient of the log-prior distribution), learnable from data \cite{song2019generative}. Differing from prior works such as \cite{Liu2021, Wang2022}, we adopt a unified network to simultaneously estimate both the direct channel and cascaded channels, independent of the number of tags engaged. This strategy streamlines our model's complexity and enhances its applicability.
 The main contributions are summarized as follows:
\begin{itemize}
    \item We present a novel method that employs an adversarial score-based generative model.  It uses a hybrid training approach to alternatively optimize adversarial and denoising score-matching objectives, enabling the learning of diverse and precise channel distributions. During the inference, we exploit the trained model to generate denoised channels through annealed sampling from the score function. 
    \item We provide empirical analyses to assess the performance of our proposed method. The proposed adversarial score-based model performs remarkably close to optimal and achieves the performance of the MMSE estimator for the direct link and outperforms it for the cascaded links, even in low signal-to-noise ratio (SNR) regimes.

\end{itemize}
Our approach is versatile and adaptable to diverse channel distributions, proving advantageous for intricate or unfamiliar distributions. This is particularly valuable when the optimal MMSE estimator cannot be implemented due to complexity or unavailability of the channel correlation matrix \cite{kay1993fundamentals}.

\noindent\textit{Notation}: The  derivative of  $f(\mathbf{X})$ with respect to  $\mathbf{X}$ is $\nabla_{{\mathbf{X}}}f(\mathbf{X}),$  $\mathcal{K} \triangleq \{1,\ldots,K\}$, and  $\mathcal{K}_0\triangleq \{0,1,\ldots,K\}$. 

\vspace{-3mm}
\section{Adversarial Score-Based Generative Models}\label{score_networks}
Score-based generative modeling aims to first train a neural network, known as noise conditional score network (NCSN), to accurately estimate the underlying data distribution and then generate new data points through sampling \cite{song2020improved, song2019generative}.
For a given set of i.i.d. samples ${\mathbf{x}_1, \ldots, \mathbf{x}_N}$ drawn from the distribution $p_{X}(\mathbf{x})$, where each sample is perturbed with varying scales of random Gaussian noise, the NCSN (denoted as $s_{\theta}(\mathbf{x})$ and parameterized by $\theta$) learns the score function of $p_{X}(\mathbf{x})$ as $ \nabla_{\mathbf{x}} \log p_{X}(\mathbf{x})$. 
After training the NCSN, the generation of new samples from $p_{X}(\mathbf{x})$ becomes feasible through only the use of this model via annealed Langevin sampling (ALS) technique \cite[Algorithm 1]{song2019generative}. This iterative procedure involves initializing the samples from an arbitrary prior distribution $\pi_{X}(\mathbf{x})$ with a step size $\beta > 0$, and then continuing sampling from the final samples of the previous distribution while gradually reducing the step size over a predetermined number of iterations $T$. 

While score-based generative models offer remarkable advantages, including the generation of highly diverse samples, the quality of the generated samples can be further improved by incorporating adversarial objectives \cite{jolicoeur2020adversarial}. The concept involves training a neural network discriminator (hereafter denoted as DiscNet) to accurately differentiate between original data and samples generated by the NCSN, which is referred to as the generator. It employs an alternating training scheme involving the discriminator and NCSN, encouraging the NCSN to generate high-quality samples with a diversity akin to that of score-based generative models (see \cite{jolicoeur2020adversarial} and reference therein).

\vspace{-3mm}
\section{System, Channel, and Signal Models}\label{system_modelA}
\subsection{System and Channel Models}
The considered system comprises a single-antenna RF source, $K$ single-antenna tags ($k$th tag is denoted by $T_k$), and a reader with $M$ antennas (Fig.~\ref{fig_SystemModel}). During each fading block, $\mathbf{h}_0 = [h_{1,0}, \ldots, h_{M,0}]^{\rm{T}} \in \mathbb{C}^{M \times 1}$ is the direct channel  vector from the RF source to the reader. Moreover, $\mathbf{h}_k = f_k \mathbf{g}_k \in \mathbb{C}^{M \times 1}$ for $k\in \mathcal{K}$ is the effective backscatter (cascaded) channel through $T_k$, which is the product of the forward-link channel from the RF source to $T_k$, i.e., $f_k \in \mathbb{C}$, and the backscatter channel from $T_k$ to the reader, i.e.,  $\mathbf{g}_k = [g_{1,k}, \ldots, g_{M,k}]^{\rm{T}}  \in \mathbb{C}^{M\times 1}$. 


In Fig. \ref{fig_SystemModel}, for each coherence block  $\tau_c$, $\tau$ ($<\tau_c$) and $\tau_c -\tau$ samples are for  channel estimation and   data transmission.

\vspace{-2.5mm}
\subsection{Tag operation}\label{EH_data} 

{For sending data and pilot, a  tag uses load modulation \cite{Rezaei2023}. It involves cycling through different impedance values  ($Z_m$) to create a multi-level signal constellation. Thus, to  generate  a symbol $c_m$ with $\mathbb{E}{\vert c_m \vert^2}=1$, the tag sets its impedance to $Z_m$ and presents it to the antenna with impedance $Z_a$, resulting in  reflection coefficient $\Gamma_m=(Z_m-Z_a^*)/(Z_m+Z_a)=\sqrt{\alpha} c_m$, where $\alpha$ represents the power reflection factor \cite{Rezaei2023}. This letter confines to constant-modulus signaling.} {Furthermore, tags have limited energy storage and transmit data and harvest energy from the RF source signal simultaneously. The harvested energy powers the tag during channel estimation (see \cite{Rezaei2023} and reference therein for more details).}

\vspace{-2mm}
\section{Channel Estimation}\label{sec:channel_estimation}
\begin{figure*}[t!]
\vspace{0pt}
\noindent\begin{minipage}[c]{0.59\textwidth}
    \centering
    \includegraphics[width=\textwidth]{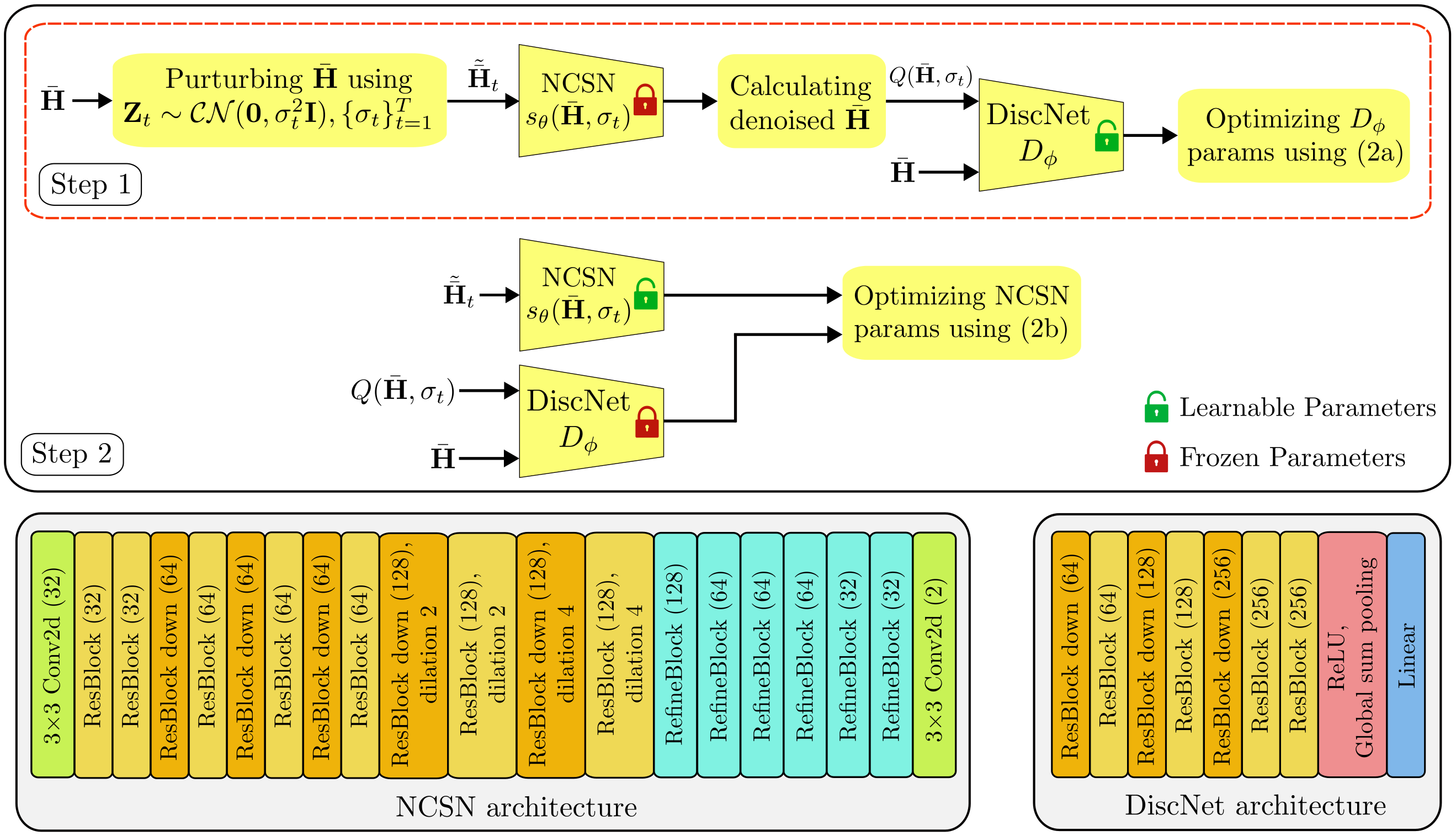}
\end{minipage}
\hfill
\begin{minipage}[c]{0.39\textwidth}
    \centering
    \begin{algorithm}[H]
    \caption{: Channel Estimation via ALS}\label{Algo1}
    {\scriptsize{\begin{algorithmic}
    \renewcommand{\algorithmicrequire}{\textbf{Require:}}
    \Require  $ \{\sigma_t \}_{t=1}^{T}, \beta_0, N, \zeta$
    \Initialize{$\hat{\bar{\mathbf{H}}}_{T}^{0} \sim \mathcal{CN}(\mathbf{0}, \sigma^2_{\rm{max}}\mathbf{I})$}
    \For{$t=T,\ldots,1$}
    
        $ \beta_t = \beta_0 \sigma_t^2/\sigma_T^2$.

         \For{$n=1,\ldots,N$}

          $\bar{\mathbf{Z}}_t^n \sim \mathcal{CN}(\mathbf{0}, \mathbf{I})$.
          
          Calculate $\nabla_{\hat{\bar{\mathbf{H}}}} {\rm{log}}~p_{\hat{\bar{H}}|Y}(\hat{\bar{\mathbf{H}}}_{t}^{n-1}|\mathbf{Y})$ using \eqref{gradian}.
          
         $ \hat{\bar{\mathbf{H}}}_t^n \leftarrow \hat{\bar{\mathbf{H}}}_{t}^{n-1}+ \beta_t \nabla_{\hat{\bar{\mathbf{H}}}} {\rm{log}}~p_{\hat{\bar{H}}|Y}(\hat{\bar{\mathbf{H}}}_{t}^{n-1}|\mathbf{Y})+ \sqrt{2\beta_t \zeta} \bar{\mathbf{Z}}_t^n$.
          \EndFor
          
        \State \textbf{return}:  $\hat{\bar{\mathbf{H}}}_{t-1}^0 = \hat{\bar{\mathbf{H}}}_{t}^N$. 
    \EndFor
\end{algorithmic}
\textbf{Output:} The channel estimate $\hat{\bar{\mathbf{H}}}$}}.
\end{algorithm}
\end{minipage}
\caption{\scriptsize{Overview of the training phase of the proposed method (best viewed in color), illustrating the NCSN as the generator network and DiscNet as the discriminator network (number of filters are shown in parenthesis). The training process involves alternating between step 1 and step 2 to learn channel distributions. During the inference phase, the trained NCSN and ALS are exclusively utilized to estimate denoised channels, following the procedure detailed in Algorithm~\ref{Algo1}.
}} \label{fig:train}
 \vspace {-0.7cm}
\end{figure*}

The goal is to estimate $\mathbf{H} = [\mathbf{h}_0, \mathbf{h}_1, \ldots, \mathbf{h}_K]$ using pilot training-based channel estimation methods. During the channel estimation phase, 
the RF source transmits a pilot sequence $\mathbf{s} =[s_1,\ldots,s_{\tau}]\in \mathbb{C}^{1 \times \tau}$, where $s_i$  satisfies $|{s}_i|^2 = 1$ for $ i = \{1.\ldots, \tau\}$\footnote{ When $\mathbf{s}$ is unknown and changes over time, blind schemes should be developed \mbox{\cite{Ye2023}}. We leave it as a future research topic.}.
 Following the methodology presented in \cite{rezaei2023timespread}, we consider that all the tags are active during the estimation interval and backscatter the RF source signal to transmit their pilot signals, i.e., $T_k$ backscatters $\mathbf{c}_k  = [c_{k1},\ldots c_{k \tau}]\in \mathbb{C}^{1 \times \tau},$ where  $c_{ki}$ is the tag's transmit pilot symbol over the $i$th RF source symbol, $s_i$. Following \cite{rezaei2023timespread}, we treat the RF source as an imaginary tag to whom an all-$\mathbf{1}$ is assigned as the pilot, i.e., $\mathbf{c}_0 = [1, \ldots, 1] \in \mathbb{R}^{1 \times \tau}$, and adopt the rows of the Hadamard matrix excluding the first row as the tags' pilots $\mathbf{c}_k, k \in \mathcal{K}$, using binary phase-shift keying (BPSK) modulation \cite{rezaei2023timespread}\footnote{Any set of mutual orthogonal sequences can be modified to be used at tags for channel estimation, e.g., modified Zadoff-Chu sequences \cite[Theorem 4]{rezaei2023timespread}.}. Hence, the first $K+1$ rows of a Hadamard matrix of order $m$, i.e., $\mathbf{H}_m^{\text{h}} \in \{1,-1\}^{m \times m}$,  are selected as  pilots, where $m = 2^q $ and $q \ge 1$, satisfying $m \ge K+1$, and $\mathbf{c}_i \mathbf{c}_j^{\rm{H}} = 0, i \ne j$ for $i,j \in \mathcal{K}_0$. Thus, for the channel estimation, $\tau = m$. The reader then estimates the direct and cascaded channels using the tags' backscattered signals and the RF source signal.

Given the above setup, the received signal at the reader over $\tau$ RF source symbol, $\mathbf{y}\in \mathbb{C}^{M \times \tau}$, is given as  \cite{rezaei2023timespread}
\begin{eqnarray}\label{received_pilot}
 \mathbf{Y} = \sqrt{p_p} \bar{\mathbf{H} } \mathbf{C} \mathbf{S} +  \mathbf{N},
\end{eqnarray}
where $p_p$ is  the pilot transmit power, $\bar{\mathbf{H}} = [\mathbf{h}_{0}, \sqrt{\alpha_1} \mathbf{h}_{1}, \ldots, \sqrt{\alpha_K} \mathbf{h}_{K}] \in  \mathbb{C}^{M \times (K+1)}$, $\mathbf{S}\triangleq \rm{diag}(\mathbf{s})$, and $\mathbf{N} \in \mathbb{C}^{M \times \tau} \mathcal{CN}(\mathbf{0}, \sigma^2\mathbf{I})$ is the noise matrix. In \eqref{received_pilot}, $\mathbf{C} = [\mathbf{c}_0, \ldots, \mathbf{c}_{K}]^{\rm{T}} \in \mathbb{C}^{(K+1) \times \tau}$, are the transmitted pilots by the imaginary tag and the other $K$ tags, and $\mathbf{C}\mathbf{C}^{\rm{H}} = \tau \mathbf{I}_{K+1}$. 


Although  {\eqref{received_pilot}} appears similar to typical multi-user pilot-based channel estimation with active radio nodes, since passive tags can only reflect external pilot symbols, the entries of   $\mathbf{\bar{H}}$ exhibit an intricate multi-modal distribution, comprising both the direct channel and the cascaded channels, each characterized by radically different fading behaviors, making accurate data distribution modeling challenging.
\vspace{-3.6mm}
\subsection{Adversarial Score-Based Channel Estimation}\label{Sec:Proposed}
In this section, we propose a DL-based method to estimate $\bar{\mathbf{H}}$ from the received signal $\mathbf{Y}$ \eqref{received_pilot}. Note that $\bar{\mathbf{H}}$ follows an unknown multi-modal distribution due to the different fading characteristics of the direct link (${\mathbf{h}}_0$) and the cascaded links $(\mathbf{h}_k, k \in {\mathcal{K}})$. 
Harnessing the power of adversarial score-based generative models to be optimized for multi-modal landscapes \cite{jolicoeur2020adversarial, song2019generative}, we propose joint estimation of ${\mathbf{h}}_0$ and $\mathbf{h}_k$ through a single network (i.e., $s_{\theta}(\mathbf{x})$) without the need for retraining for each channel.
The proposed method comprises two phases: i) training the NCSN and DiscNet using an adversarial score-matching approach on noise-perturbed channel distributions (Fig.~\ref{fig:train}), and ii) jointly estimating the direct and cascaded links of the channel using the ALS technique (Algorithm~\ref{Algo1}).
 The training process is performed offline by using a small dataset (training samples) of channel measurements or simulated channel realizations. It aims to model the channel distribution $\bar{\mathbf{H}}$ by estimating the score function, i.e., the gradient of the log probability density with respect to data. During the testing phase, our proposed method uses the trained model and online observation $\mathbf{Y}$ (1) to jointly estimate the direct and cascaded channels via the ALS technique \mbox{\cite{song2019generative}}. %

During the training phase, the NCSN learns to estimate the score function of the logarithmic density of the  channels, while the DiscNet is sequentially trained to distinguish the probability of denoised channels from that of original data (ensuring high quality and diversity in our generative model). 
In the inference (or testing) phase, we employ the ALS technique for channel estimation solely using the trained NCSN model. 
The training phase does not rely on the information of the pilot signals, $\mathbf{S}$ and $\mathbf{C}$ \eqref{received_pilot}, and the noise power, $\sigma^2$, making the inference phase robust and adaptable across a wide range of SNR values and a number of pilot sequences. 

\subsubsection{\textbf{Learning Channel Distributions}}
Consider a sequence of positive noise scales $\{\sigma_t\}_{t=1}^T$ that satisfy $\sigma_{\rm{min}}=\sigma_1<\sigma_2< \ldots <\sigma_T = \sigma_{\rm{max}}$\footnote{This set of noise variances is defined as a geometric progression between $\sigma_1$ and $\sigma_T$, with $T$ chosen according to some computational budget \cite{jolicoeur2020adversarial}.}. 
To facilitate exploration of the channel distribution in both low-density and high-density regions, we first perturb each channel sample $\bar{\mathbf{H}}$ with the complex Gaussian noise $\mathbf{Z}_t \sim \mathcal{CN}(\mathbf{0}, \sigma_t^2 \mathbf{I})$, where $\sigma_t \in \{\sigma_t\}_{t=1}^T$.
As a result, we obtain a noise-perturbed sample $\tilde{\bar{\mathbf{H}}}_t = \bar{\mathbf{H}} + {\mathbf{Z}}_t$, where $p_{\tilde{\bar{{H}}}_t|\bar{{H}}}(\tilde{\bar{\mathbf{H}}}_t|\bar{\mathbf{H}})\sim \mathcal{CN}(\bar{\mathbf{H}}, \sigma_t^2 \mathbf{I})$.  
Typically, $\sigma_{\rm{min}}$ is chosen to be small enough such that $p_{\tilde{\bar{{H}}}_t}(\tilde{\bar{\mathbf{H}}}_t) = p_{{\bar{{H}}}_t}({\bar{\mathbf{H}}}_t)$, while $\sigma_{\rm{max}}$ is selected to be sufficiently large such that $ p_{\tilde{\bar{{H}}}_t}(\tilde{\bar{\mathbf{H}}}_t) \sim \mathcal{CN}(\mathbf{0}, \sigma^2_{\rm{max}}\mathbf{I})$ \cite{song2019generative}.

\begin{figure*}
\vspace{-0.5mm}
\begin{subequations} \label{theta_star_total}
\begin{eqnarray}
         && \max_{\phi}  \mathbb{E}_{p_{\bar{H}}(\bar{\mathbf{H}})}  \left\{ \left(D_{\phi}(\bar{\mathbf{H}})-1\right)^2 \right\}+ \mathbb{E}_{p_{\bar{H}}(\bar{\mathbf{H}})}  \mathbb{E}_{p_{{\tilde{\bar{{H}}}_t|\bar{{H}}}}(\tilde{\bar{\mathbf{H}}}_t|\bar{\mathbf{H}})} \left\{ \left(D_{\phi} (Q(\bar{\mathbf{H}}, \sigma_t))+1 \right)^2  \right \}\label{theta_star1}\\
   &&  \min_{\theta} 
 \mathbb{E}_{p_{\bar{H}}(\bar{\mathbf{H}})}  \mathbb{E}_{p_{{\tilde{\bar{{H}}}_t|\bar{{H}}}}(\tilde{\bar{\mathbf{H}}}_t|\bar{\mathbf{H}})} \left\{ \left(D_{\phi} (Q(\bar{\mathbf{H}}, \sigma_t))-1 \right)^2  + \frac{\lambda}{2}\sigma_t^2 \left\Vert  s_{\theta}(\bar{\mathbf{H}}, \sigma_t) -\nabla_{\tilde{\bar{\mathbf{H}}}}  p_{\tilde{\bar{{H}}}_t|\bar{{H}}}(\tilde{\bar{\mathbf{H}}}_t|\bar{\mathbf{H}}) \right \Vert^2 \right \}.\label{theta_star} 
\end{eqnarray}
\end{subequations}

\vspace{-3mm}
\hrulefill
\vspace{-3.8mm}
\end{figure*}

As shown in Fig.~\ref{fig:train}, we train the NCSN, denoted as $s_{\theta}(\tilde{\bar{\mathbf{H}}}_t) = s_{\theta}(\bar{\mathbf{H}}, \sigma_t)$, to learn the score of the conditional distribution $p_{\tilde{\bar{{H}}}_t|\bar{{H}}}(\tilde{\bar{\mathbf{H}}}_t|\bar{\mathbf{H}})$, incorporating the perturbed channel $\tilde{\bar{\mathbf{H}}}_t$. 
We adopt the hybrid adversarial approach proposed in \cite{jolicoeur2020adversarial}, alternately minimizing the score-matching loss for the NCSN and maximizing the adversarial loss for the DiscNet with the objective given in \eqref{theta_star_total}. During the training phase, the NCSN attempts to estimate an uncorrupted channel from a noisy input channel by minimizing the $l_2$ distance between them. 
On the other hand, the DiscNet strives to increase the similarity between the distribution of the original channel $\bar{\mathbf{H}}$ and the distribution of the generated (denoised) channel. It thus encourages the NCSN to generate a denoised channel that is more realistic from the perspective of the DiscNet. 
In the first step, we freeze the NCSN and train the DiscNet using the least square GAN (LSGAN) formulation \eqref{theta_star1}, in which $Q(\bar{\mathbf{H}}, \sigma_t) = s_{\theta}(\bar{\mathbf{H}}, \sigma_t)\sigma^2_t + \tilde{\bar{{H}}}_t$ represents the recovered denoised channel through the score function employing the Empirical Bayes mean \cite{jolicoeur2020adversarial}. 
In the second step, we freeze the DiscNet and proceed to train the NCSN using the adversarial objective function \eqref{theta_star}. The second term within this objective corresponds to the weighted denoising score matching objective \cite{song2019generative}, and it can be further simplified by substituting $\nabla_{\tilde{\bar{\mathbf{H}}}}  p_{\tilde{\bar{{H}}}_t|\bar{{H}}}(\tilde{\bar{\mathbf{H}}}_t|\bar{\mathbf{H}})  = -\mathbf{Z}_t/\sigma^2_t$. 
Here, $\lambda$ refers to a hyperparameter that regulates the relative influence of the denoising score-matching objective and the adversarial loss.
Note that all expectations in \eqref{theta_star_total} can be efficiently estimated using empirical averages \cite{song2020improved}. The training phase involves alternatively applying these two steps until convergence (Fig.~\ref{fig:train}).

\subsubsection{\textbf{Channel Estimation via ALS}}
After training, we solely employ the trained NCSN and ALS technique to estimate $\bar{\mathbf{H}}$ during the inference phase. Initially, the ALS employs scores associated with the highest noise level and progressively anneals down the scale until it reaches a point where it cannot be differentiated from the original channel distribution.
Given $\mathbf{Y}$ \eqref{received_pilot}, we apply $N$ steps of ALS to sample from the posterior distribution $ p_{\bar{{H}}|{Y}}(\bar{\mathbf{H}}|\mathbf{Y})$ \cite{song2019generative}. The channel estimation at the $n$-th step is thus obtained as
\begin{eqnarray}\label{eqn:update}
  \hat{\bar{\mathbf{H}}}_t^n \leftarrow \hat{\bar{\mathbf{H}}}_{t}^{n-1}+ \beta_t \nabla_{\hat{\bar{\mathbf{H}}}} {\rm{log}}~p_{\hat{\bar{H}}|Y}(\hat{\bar{\mathbf{H}}}_{t}^{n-1}|\mathbf{Y})+ \sqrt{2\beta_t \zeta} \bar{\mathbf{Z}}_t^n,~
\end{eqnarray}
for $1 \le n \le N$, and $\sigma_t \in \{\sigma_t \}_{t=1}^{T}$. In \eqref{eqn:update}, $\bar{\mathbf{Z}}_t^n \sim \mathcal{CN}(\mathbf{0}, \mathbf{I})$ is added at every sampling step. 
The step size $\beta_t = \beta_0 {\sigma_t}^2/{\sigma_T}^2$, where $\beta_0$ and $\zeta$ represent the initial step size and the scale factor for sample diversity, respectively \cite{jalal2021instance}. These values will be determined through the grid search \cite{gridsearch}. 
To compute the second term of \eqref{eqn:update}, we apply the Bayesian rule as given by ${\rm{log}}~  p_{\bar{{H}}|{Y}}(\hat{\bar{\mathbf{H}}}_{t}^{n-1}|\mathbf{Y}) = {\rm{log}}~  p_{{Y}|\bar{{H}}}(\mathbf{Y}|\hat{\bar{\mathbf{H}}}_{t}^{n-1}) + {\rm{log}}~  p_{\bar{{H}}}(\hat{\bar{\mathbf{H}}}_{t}^{n-1})- {\rm{log}}~ p_{Y}(\mathbf{Y}).$
The  gradient with respect to $\hat{\bar{\mathbf{H}}}$ is as follows:
\begin{eqnarray}\label{gradian}
 \nonumber \nabla_{\hat{\bar{\mathbf{H}}}} {\rm{log}}~  p_{\hat{\bar{{H}}}|{Y}}(\hat{\bar{\mathbf{H}}}_{t}^{n-1}|\mathbf{Y})  &=&  -\frac{(\mathbf{Y} - \sqrt{p_p}\hat{\bar{\mathbf{H}}}_{t}^{n-1}\mathbf{C} \mathbf{S})\sqrt{p_p} \mathbf{S}^{\rm{H}}\mathbf{C}^{\rm{H}}}{\sigma^2} \\&+&\underbrace{\nabla_{\hat{\bar{\mathbf{H}}}}{\rm{log}}~  p_{\hat{\bar{{H}}}}(\hat{\bar{\mathbf{H}}}_{t}^{n-1})}_{s_{\theta}(\hat{\bar{\mathbf{H}}}, \sigma_t)}.
\end{eqnarray}
 Here, $\nabla_{\hat{\bar{\mathbf{H}}}}{\rm{log}}~ p_{Y}(\mathbf{Y}) = 0$, and $\nabla_{\hat{\bar{\mathbf{H}}}}{\rm{log}}~  p_{{Y}|\hat{\bar{{H}}}}(\mathbf{Y}|\hat{\bar{\mathbf{H}}}_{t}^{n-1})$ is determined using the property that $p_{{Y}|\hat{\bar{{H}}}}(\mathbf{Y}|\hat{\bar{\mathbf{H}}}_{t}^{n-1})\sim\mathcal{CN}(\sqrt{p_p}\hat{\bar{\mathbf{H}}}_{t}^{n-1}\mathbf{C} \mathbf{S}, \sigma^2 \mathbf{I})$ \eqref{received_pilot}. Accordingly, the channel estimation \eqref{eqn:update} is feasible by having $s_{\theta}(\hat{\bar{\mathbf{H}}}, \sigma_t)$ (i.e., trained NCSN) and the sampling process continues iteratively for $T, T-1, \ldots, 1,$ where $\hat{\bar{\mathbf{H}}}_T^{0} \sim \mathcal{CN}(\mathbf{0},\sigma^2_{\mathrm{max}}\mathbf{I})$, and $\hat{\bar{\mathbf{H}}}_t^{0} = \hat{\bar{\mathbf{H}}}_{t+1}^{N}$, for $n<N$. Detailed steps of the channel estimation process using ALS are summarized in Algorithm \ref{Algo1}.

\begin{table}[hbt!]
\caption{ {Simulation settings.}} \vspace{-2mm}
\label{tab:sim_para}
\centering 
\begin{tabular}{c c c c c c} 
\hline 
Parameter & Value & Parameter & Value & Parameter & Value \\ [0.5ex] 
\hline \hline
$K, \tau$ & 7, 8 & $M$ & 48 &$T$ & 2311\\
$N$ & 6 & $\alpha$ &  0.6 & $\beta_0$ & $3 \times 10^{-9}$  \\
$\sigma^2_{\rm{max}}$&  36.77& $\sigma^2$ & 1 & $\zeta$ & $10^{-4}$\\
\hline 
\end{tabular}
\label{table:nonlin} 
\vspace{-5.5mm}
\end{table}

\section{Simulation Results}\label{sim}
Herein, we evaluate the proposed channel estimation scheme and several prior arts.

\textbf{\textit{Parameter Settings and Definitions}}:
We consider  $K=7$ tags. All channels, i.e., $\mathbf{h}_0, \mathbf{g}_k, f_k, \forall k$, are assumed to be independent quasi-static Rayleigh fading, with $\sigma^2 = 1$. We set $\tau = 8$, and $\alpha_k = 0.6, \forall k$. 

The employed unconditional NCSN and DiscNet architectures are based on \cite{song2020improved} and \cite{jolicoeur2020adversarial}, respectively. The networks include ResBlock (a residual block to extract intricate features from the data), ResBlock down (a downsampling residual block to facilitate efficient processing), ResBlock down dilation (a dilated downsampling residual block to capture a broader contextual overview), RefineBlock (a multi-path refinement block to produce precise predictions), Conv2d (2D convolutional layer), ReLU (rectified linear unit activation function that introduces non-linearity into the model), Global sum pooling (a pooling operation to capture global information), and Linear layer (a fully connected layer to perform a linear transformation) - see Fig. \ref{fig:train}. Details of the network designs and their layers can be found in \cite{song2020improved, BigGAN}. Hyperparameters were configured according to Table~\ref{tab:sim_para}. Our networks were implemented using PyTorch and trained on 10,000 samples with a batch size of 32 for 600 epochs on an Nvidia Tesla V100 GPU with 16GB RAM.

 \begin{figure*}
 \vspace{-4mm}
 \centering
 \begin{minipage}{.29\textwidth}
   \centering
   \includegraphics[width=0.9\columnwidth]{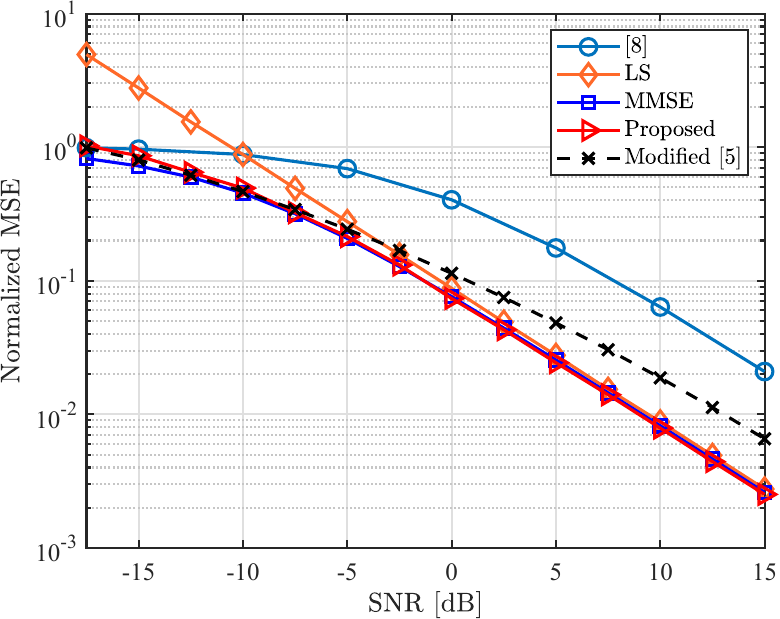}\vspace{-1.75mm}
    \caption{{NMSE versus SNR for the direct channel.}}
	\label{fig:direct_channel} \vspace{-0mm}
  \end{minipage}\hspace{3mm}
 \begin{minipage}{.29\textwidth}
   \centering
   \includegraphics[width=0.9\columnwidth]{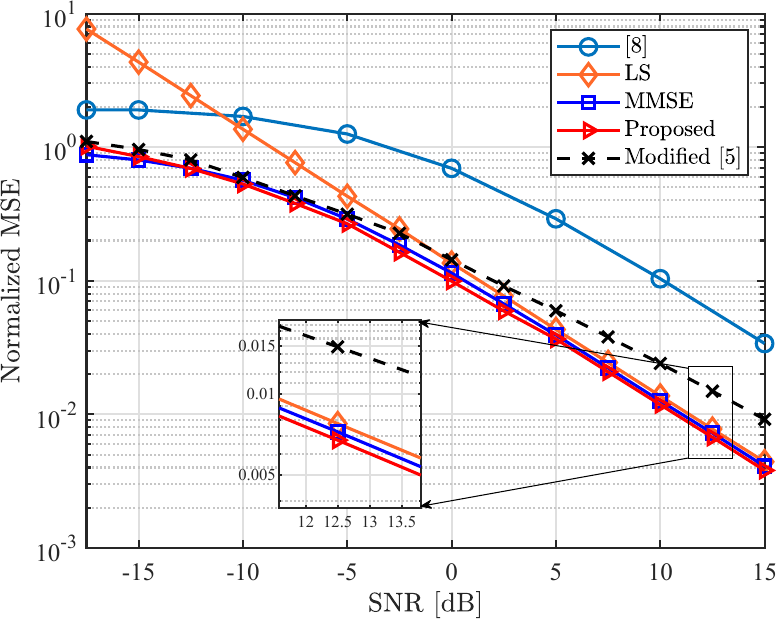}\vspace{-1.75mm}
    \caption{{NMSE versus SNR for the direct channel.}}
	\label{fig:cascade_channel} 
  \end{minipage}\hspace{3mm}
  \begin{minipage}{.29\textwidth}
   \centering
   \includegraphics[width=0.9\columnwidth]{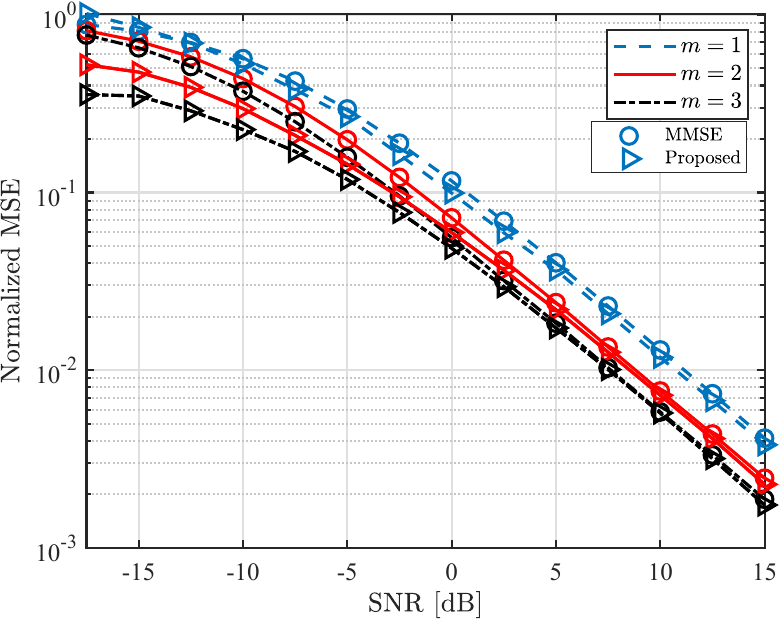}\vspace{-1.75mm}
    \caption{{NMSE versus SNR for the cascaded channel.}}
	\label{fig:distribution}
  \end{minipage}
  \vspace{-5.6mm}
 \end{figure*}
\textbf{\textit{Simulation Analysis}}:
We analyze three comparative benchmarks, including both classical estimators (LS and MMSE estimators \cite[Section III-E]{rezaei2023timespread}) and the residual deep learning-based estimator \cite{Liu2021}. To derive the LS and MMSE estimators, $\mathbf{Y}' = \mathbf{Y}\mathbf{S}^{\rm{H}}$ \eqref{received_pilot} is used. Reference  \cite{Liu2021} also estimates the direct and the cascaded channels sequentially, adopting the constraint that only one tag reflects pilot sequence, while all other tags remain silent. This method explicitly incorporates the noise variance and pilot sequences during training, necessitating the need to train the network for each SNR value individually.

Since  \mbox{\cite{rezaei2023timespread}} is the only prior study on joint direct and cascaded {\abc} channel estimation, we use it for comparative assessments. As well, we have adopted a score-based approach in \mbox{\cite{Arvinte2023}} for multi-user multi-input multi-output (MIMO) channel estimation to our problem at hand.

The quality of channel estimators is  assessed in terms of normalized MSE, which  is defined as $\text{Normalized MSE}_k = \mathbb{E} \{ { \Vert  \mathbf{h}_k - \hat{\mathbf{h}}_k  \Vert^2_2} /{\Vert \mathbf{h}_k  \Vert^2_2 }\}, \quad  k \in {\mathcal{K}_0},$
where $ \hat{\mathbf{h}}_k$ is the $k$th column of estimated channel matrix $\hat{\bar{\mathbf{H}}}$.

Fig. \ref{fig:direct_channel} and Fig. \ref{fig:cascade_channel} respectively show the NMSE performance of different estimators versus the SNR for the direct and cascaded channels.  Our method accurately estimates the multi-modal channel distribution and demonstrates remarkable accuracy in multiple tag scenarios. It significantly outperforms the LS estimate and delivers comparable performance to the optimal MMSE estimator.  In particular, our method achieves a SNR gain of $\sim$ \qty{2.5}{\dB} for an NMSE of $10^{-0.6}$ compared with the LS method for both the direct and cascaded channels. This is because, unlike the LS method which treats the channel coefficients as deterministic but unknown constants, the proposed and MMSE methods treat the channel coefficient as random with prior PDFs. Thus, these two estimators can exploit the prior statistical knowledge of the channel matrices to improve the estimation accuracy. Our method also slightly outperforms the
 implemented optimal MMSE for the cascaded channel\footnote{Since the distribution of the cascaded channels is intractable in general, the optimal MMSE estimator \cite[Section 11.4]{kay1993fundamentals} is implemented by assuming an approximate distribution \cite{rezaei2023timespread}}.. 
 As observed, the Modified method \cite{Arvinte2023} falls short in accurately estimating the direct and cascaded links. In contrast, our method excels at effectively estimating the multi-modal distribution, thereby producing high-quality samples for both the direct and cascaded channels. Notably, for the cascaded channels, our method achieves a substantial SNR improvement of $\sim$  {\qty{4}{\dB}} for an NMSE of $10^{-2}$ when compared to modified \cite{Arvinte2023}.

In contrast, \cite{Liu2021}  estimates the channels one-by-one and  uses a fraction of the pilot sequence for each channel (one pilot symbol per channel for $\tau = 8$ and $K=7$), whereas the other methods including ours estimate the channels simultaneously,  utilizing the entire pilot length for each channel, leading to  higher accuracy. Specifically, at an NMSE of $10^{-1},$  the SNR gain of  joint estimation is $\sim$ \qty{8.5}{\dB} and $\sim$ \qty{10}{\dB} for  the direct  and cascaded channels, respectively.

Fig. {\ref{fig:distribution}} also explores our approach's performance across different channel distributions, i.e., Nakagami-$m$ with $m \in \{1,2,3\}$,  where $m =1$ represents Rayleigh fading. Our method consistently outperforms MMSE for the cascaded channels, with more significant gains as $m$ increases.



\vspace{-4mm}
\section{Conclusion}
This letter introduces a novel  \abc channel estimation method through adversarial score-based generative models. Our approach yields precise channel distribution estimations, yielding remarkably high accuracy and outperforming previous methods. It matches the performance of the MMSE estimator for the direct channel and surpasses the latter for the cascaded channels without relying on channel statistics.

Future research offers numerous directions. Our channel estimates can be exploited not only for detection and decoding but also for various critical communication tasks. Additionally, exploring hardware imperfections and nonlinear distortions, as highlighted in \mbox{\cite{Qing2022, Ye2023}}, can significantly enhance our understanding. Additionally, the current focus on reflecting intelligent surfaces (RIS) and relay-assisted communications, both involving cascaded channels, aligns well with the potential solutions our method can offer in these domains.

\vspace{-0.3cm}

\linespread{1.0}
\bibliographystyle{IEEEtran}
\bibliography{IEEEabrv,Ref}

\end{document}